\documentclass[sigplan,screen,nonacm]{acmart}

\newif\ifpreprint

\preprinttrue

\acmConference[Arxiv]{}{Preprint}{To appear at PLDI}
\acmYear{2022}
\acmISBN{} 
\acmDOI{} 
\startPage{1}
\setcopyright{none}
\usepackage{booktabs}   
\usepackage{subcaption} 
\usepackage{xspace, listings, xcolor}

\undef \comment
\usepackage[final]{changes}

\newcommand{\R}{\v{R}\xspace}

\definecolor{Pallete1}{RGB}{27,147,119}
\definecolor{Pallete2}{RGB}{217,95,2}
\definecolor{Pallete3}{RGB}{117,112,179}
\definecolor{Pallete4}{RGB}{27, 158, 119}
\definecolor{Pallete5}{RGB}{217, 95, 2}
\definecolor{Pallete6}{RGB}{117, 112, 179}
\definecolor{Pallete7}{RGB}{102,102,102}

\definecolor{LightGray}{rgb}{.92,.92,.92}
\definecolor{Gray}{rgb}{.3,.3,.3}
\definecolor{DarkGray}{rgb}{.5,.5,.5}

\newcommand{\eg}{\emph{e.g.,}\xspace}
\newcommand{\ie}{\emph{i.e.,}\xspace}

\renewcommand{\c}[1]{{\lstinline!#1!}\xspace}

\lstdefinestyle{R}{ %
  language=R,
  deletekeywords={data, length, warning, env, equal, runif, trace, readline, args, exp, t, all, <, -, add,
                  warnings, max, system, time,list,c,rm,sys,frame,any,is,na, length, double},
  morekeywords={+},
  otherkeywords={<-,<<-,!,!=,~,\$,*,\&,+,^,\%,\%/\%,\%*\%,\%\%,>,/, = },
}
\lstdefinestyle{C}{ %
  language=C++,
  morekeywords={Value},
}
\lstdefinestyle{pir}{ %
  morekeywords={Checkpoint, D1},
}
\lstdefinestyle{llvm}{ %
  morekeywords={osr, cont},
}

\begin{document}

\newcommand{\bigDataframeNumberOfColumns}{50\xspace}
\newcommand{\bigDataframeNumberOfRows}{10^7\xspace}
\newcommand{\bigDataframeNumberOfExecutions}{10\xspace}

\newcommand{\bigDataframeBaselineTimePerIterationStabilizedSeconds}{0.39\xspace}
\newcommand{\bigDataframeBaselineTimePeakPerformanceSeconds}{0.011\xspace}

\newcommand{\bigDataframeDeoptlessTimeExecutionAndCompilationDropToSeconds}{0.045\xspace}

\newcommand{\bigDataframeDeoptlessImprovementAgainstBaseline}{$35\times$\xspace}

\newcommand{\perfMed}{1.9}
\newcommand{\perfMin}{1}
\newcommand{\perfMax}{9.1}
\newcommand{\memMed}{4}
\newcommand{\memMin}{22}
\newcommand{\memMax}{45}

\title[Deoptless]{Deoptless: Speculation with
  Dispatched On-Stack Replacement and Specialized Continuations}

\author{Olivier Fl\"uckiger}
  \affiliation{\institution{Northeastern University}}
  \orcid{0000-0003-0512-9607}
\author{Jan Je\v{c}men}
  \affiliation{\institution{Czech Technical University}}
  \orcid{0000-0002-4543-4262}
\author{Sebastián Krynski}
  \affiliation{\institution{Czech Technical University}}
  \orcid{0000-0002-4124-0225}
\author{Jan Vitek}
  \affiliation{\institution{Northeastern University}}
  \affiliation{\institution{Czech Technical University}}
  \orcid{0000-0003-4052-3458}

\begin{abstract}
  Just-in-time compilation provides significant performance improvements for programs written in
  dynamic languages. These benefits come from the ability of the compiler to speculate about
  likely cases and generate optimized code for these. Unavoidably, speculations
  sometimes fail and the
  optimizations must be reverted. In some pathological cases, this can leave the program stuck with
  suboptimal code. In this paper we propose \emph{deoptless}, a technique that replaces deoptimization
  points with dispatched specialized continuations. The goal of deoptless is to take a step
  towards providing users with a more transparent performance model in which mysterious slowdowns
  are less frequent and grave.
\end{abstract}

\maketitle

\section{Introduction}

At the heart of many high-performance just-in-time compilation strategies lies the ability to
replace code while it is being executed. A typical two-tier architecture has an interpreter for
quick startup, and a compiler for peak performance. In this architecture, when an implementation
needs to tier-up as it is evaluating a long-running function, one should not need to wait for the
interpreter to complete, but rather switch immediately to a natively compiled version~\citep{hol94}.
Or, when speculative compilation is found to be wrong, execution must not continue and the current
code must be replaced with a correct version~\citep{hol92}. In yet another example, the compilation of
unlikely code paths can be deferred~\citep{cha91}.

\begin{figure}[t]
  \includegraphics[width=0.85\columnwidth]{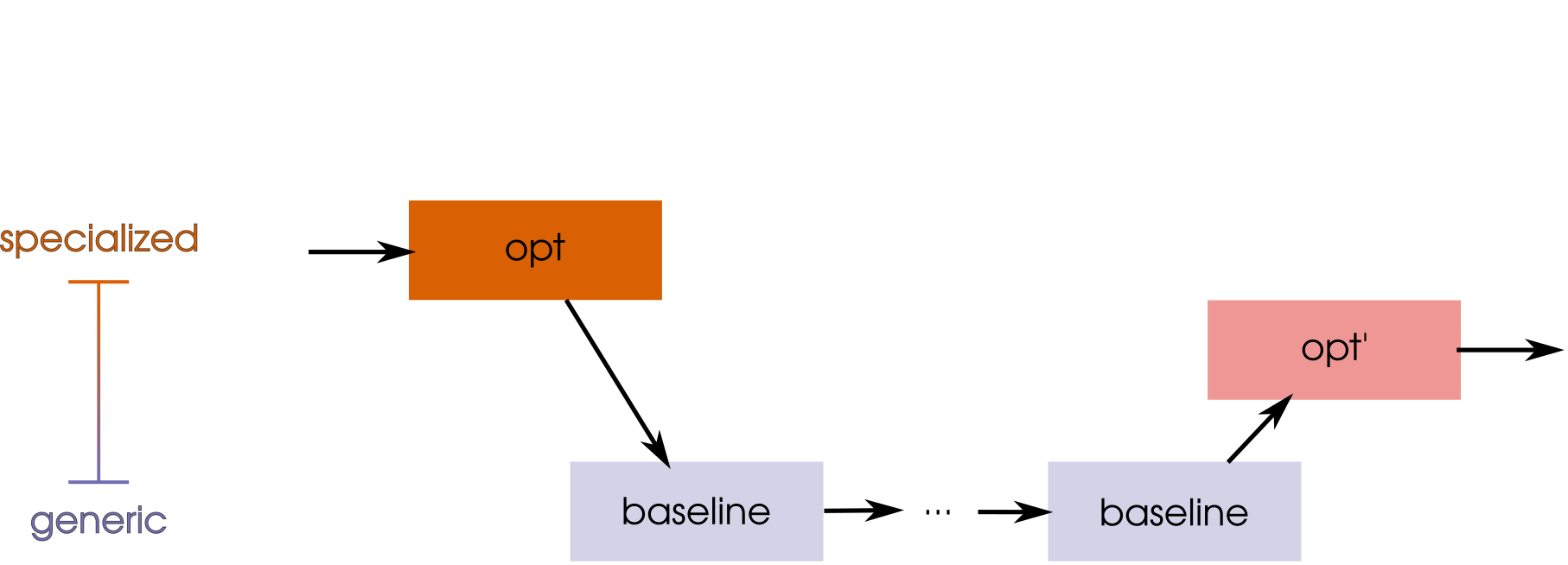}
  \caption{Deoptimization: OSR-out, profile, recompile}
  \label{fig:compare-deopt}
  \bigskip
  \ifpreprint\bigskip\else\fi
  \includegraphics[width=0.85\columnwidth]{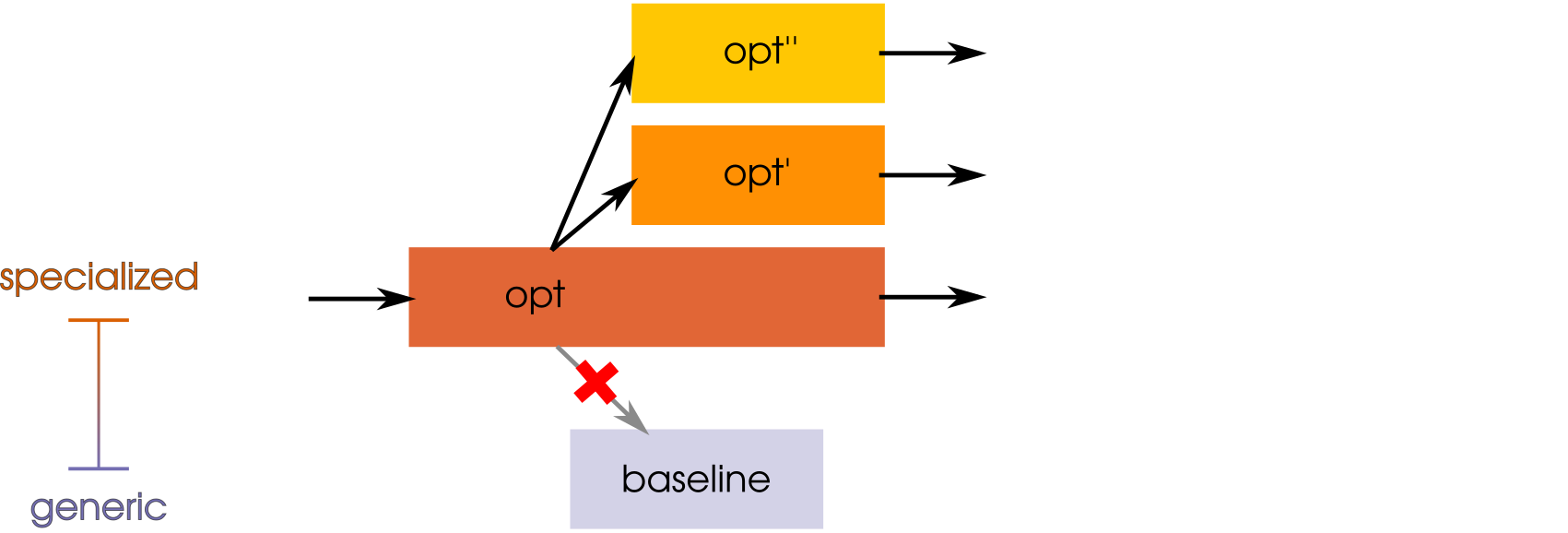}
  \caption{Deoptless: dispatched OSR to specialization}
  \label{fig:compare-deoptless}
  \Description[Control-flow-graph for deoptimization and
  deoptless]{Abstracted control-flow-graph, showing deoptimization tiering down
  in contrast to deoptless specializing.}
  \ifpreprint\bigskip\fi
\end{figure}

The common theme is replacing code with currently active stack frames, hence the name
\emph{on-stack replacement} (OSR). This may refer to pieces of code that have the same format, \eg
replacing native code with native code at a different optimization level, or a completely different
format, \eg switching from native code to interpreted code.
From a distance, OSR can be described as a mechanism for suspending execution of a function, rewriting
its state, and resuming execution in a different version of that function. What makes OSR challenging
is the low-level interaction with execution states and the need to establish a mapping between versions
of a function. For example, consider a function in which some variable was constant-folded away.
In order to transfer control to a version of the same function in which this optimization was not
applied, one must first establish a correspondence between program counters, then reconstruct the
value of the variable, and store it at the expected position among the function's locals.

Speculative compilation is characterized by code compiled under assumptions.
Any of these assumptions may prove to be false at runtime, and thus the compiler inserts
guards which trigger OSR to avoid executing miscompiled code. This situation
is often informally referred to as \emph{deoptimization} or as OSR-out
(as OSR is used to exit code).
To illustrate, consider a function that operates on a list of numbers. At
run-time, the system observes the type of the values stored in the list.
After a number of calls, if the compiler
determines that the list holds only integers, it will speculate that this will remain true
and generate code optimized for integer arithmetic.
If, at some later point, floating point numbers appear instead, a deoptimization will
be triggered.  As shown in \autoref{fig:compare-deopt}, OSR-out makes it
possible to swap the
optimized code in-place with the baseline version of the function. In subsequent calls to the
function the compiler refines its profiling information to indicate that the function
can operate on lists of integers and floating point numbers. Eventually, the function will
be recompiled to a new version that is slightly more general. That version will not
need to deoptimize for floating point values, but likely will not be as efficient as the
previously optimized one.

Speculative compilation can cause hard to predict performance pathologies. Failed
speculations lead to two kinds of issues.
First, deoptimization causes execution to suddenly slow down as the new code being executed
does not benefit from the same level of optimization as before.
Second, to avoid repeated deoptimizations, the program eventually converges
to code that is more generic, \ie that can handle the common denominator of all
observed executions. From a user's point of view, the program speeds up again,
but it does not regain its previous performance.

In this paper we present \emph{deoptless}, a strategy for avoiding
deoptimization to a slower tier. The idea is to handle failing assumptions
with an optimized-to-optimized transfer of control.
At each deoptimization point, the compiler maintains multiple optimized continuations,
each specialized under different assumptions. When OSR is triggered, 
a continuation that best fits the current state of execution is selected.
The function that triggered OSR is also not retired with deoptless 
(as would occur in the normal case), rather it is retained in the hope that it can be used again.

\autoref{fig:compare-deoptless} illustrates what happens when speculation fails with deoptless.
Instead of going to the baseline, the compiler generates code for the continuation, and
execution continues there. This can result in orders-of-magnitude faster
recovery from failed speculation.
Furthermore, deoptless not only avoids tiering down, it also gives the compiler an
opportunity to generate code that is specific to the current execution context.
As we later demonstrate, this can significantly increase the peak performance of
generated code.
For instance, if an assumption fails, as above, because a list holds
floating point numbers rather than integers, then the continuation can be specialized to
handle floats. In subsequent executions, if the same OSR point is reached,
the continuation to invoke will be selected by using \emph{context dispatch}~\citep{oopsla20c}.
If no previously compiled continuation matches the execution context, then a new one
will be compiled. Of course, the number of continuations is bounded, and when that
bound is reached deoptless will deoptimize.

The contribution of this paper is the description of deoptless and an evaluation in the context of \R, a just-in-time compiler
for the R language.
We start with a background on OSR in~\autoref{sec:osr}.
This is followed by a description of the deoptless compilation
strategy in \autoref{sec:deoptless}, and details of our prototype implementation
in \autoref{sec:rsh-deoptless}.  The performance evaluation described in \autoref{sec:eval}
shows how much faster deoptless can handle failing assumptions on average and
some of the potential performance gains.

One major limitation, which we are upfront about, is that our performance
evaluation is limited to synthetic benchmarks. \R is a research compiler, and while it is able
to run all R programs, for most real-world applications it still has a ways to go.
Moreover, in programs where \R shows good performance, typically few deoptimizations happen.
The drawback of this state of affairs is that we are not able to quantify how often
the performance pathologies we are targeting occur in practice. It is our belief
that they will occur and that they have the potential to be significant.

\added[comment=unblind]{
\R as well as the presented contributions are freely available at \href{https://r-vm.net}{ř-vm.net} and archived as an artifact to reproduce the experimental section~\citep{flu22-artifact}.
}

\section{Background: On-Stack Replacement}\label{sec:osr}

On-stack replacement (OSR) refers to an exceptional transfer of control
between two versions of a function. It is employed by just-in-time compilers in situations
where a function can or has to be replaced at once, without waiting for it
to exit normally. To the user, this exchange is not observable, the new function
transparently picks up where the old one stopped. \emph{On-stack} refers to
the fact that the involved functions have active stack frames that need to be
rewritten.

\paragraph{Definitions and Models}

We call functions that should be exited \emph{origins} and their
replacements \emph{targets}. Each function has an \emph{execution state}, or
\emph{stack frame}, that is dependent
\begin{figure}
  \includegraphics[width=0.95\columnwidth]{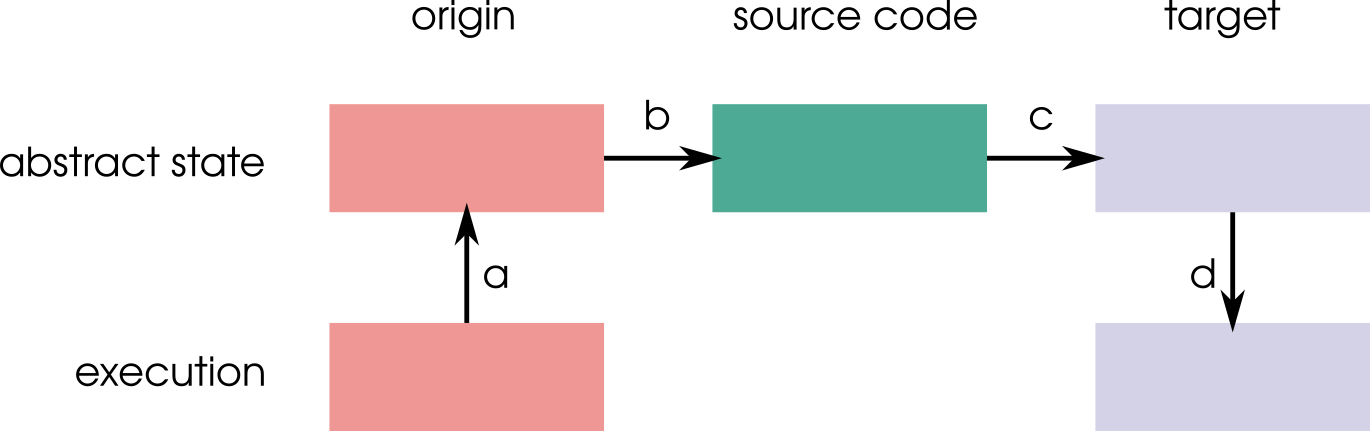}
  \caption{Parts of an OSR event}
  \Description[Abstract control-flow for OSR]{Showing the four transitions, from
  execution state to abstract state, to target abstract state and finally to
  target execution state.}
  \label{fig:osr}
  \ifpreprint\bigskip\fi
\end{figure}
on the code format but typically consists of at least the position in the code
and the values of local variables. The format of origin and target can be vastly
different if, for instance, one of them is interpreted and the other runs natively.
A \emph{mapping} between states captures the steps needed to rewrite origin
states to target states. Since both origin and target are derived from the same source code,
we sometimes use the term \emph{source} to refer to the common ancestry of various
compiled code fragments. \autoref{fig:osr} shows an idealized OSR that (a) extracts
the state of the origin, (b) maps it to the source, (c) maps it
to the target, and finally (d) materializes the target state.
Origin and target do not need to be constrained to a single stack frame and a
single function. For example when exiting an inlined function, one origin
function maps to multiple target functions. In other words, the stack frame of the origin needs to be split into
multiple target stack frames.

In practice, many implementations follow a simplified design combining (b) and
(c) into one mapping that translates directly from one state to another. This 
works because the compiler of one end of the OSR uses the
code of the other end of the OSR as source code, rather than the actual source, \eg typically the
bytecode is the source code for the optimizing native compiler:
\[
  source \rightarrow BC \rightarrow native
\]
In this architecture, there is only one compiler and one compilation direction
between the two ends of the OSR, therefore the mapping takes just one step.

On the other hand, in the case where both ends of the OSR are compiled from some
common source code, the mapping of execution states has two steps. The
first compilation defines a mapping that lifts the state from an origin state to
a source state, the second compilation a mapping that lowers it to a target
state. Therefore, the generic model is important in cases where
OSR transitions from optimized to optimized code. This was also
noted by \citet{wim17}, who describe it as ``a two-way matching of two scope
descriptors describing the same abstract frame.''
The one-step architecture is further simplified in optimizers with identical
source and target language, in which case
the states on both ends of OSR have the same
representation~\citep{ber16,wan18,popl21}.

If OSR jumps from optimized to unoptimized code, we call it \emph{OSR-out}; or
\emph{deoptimization}, when it is used to bail out of failing speculative optimizations.
If it jumps from unoptimized to optimized code, we call
it \emph{OSR-in} or \emph{tiering up}. This is useful, for instance, when the program is stuck
in a long-running loop. In the general case where it
jumps from optimized to optimized code, both apply and we simply
call it OSR. Typically OSR cannot happen at arbitrary locations; we call the
possible locations \emph{OSR exit} or \emph{OSR entry} points. \citet{popl18} showed
that placing OSR exits after every observable effect allows for OSR exits at arbitrary
locations modulo code-motion.

\paragraph{Speculative Optimizations}

Consider, as a simple example, using OSR to undo constant-folding to
support debugging. When the debugger is attached to the program,
execution is paused and the program counter is moved from the current
optimized function to the equivalent location in a version of the same function
without the constant-folding applied. Any variables removed by
constant-folding are recreated from metadata.
OSR is general as it allows undoing arbitrary transformations. When
OSR is used to transition between different optimization levels, it
must be transparent, \ie OSR becomes part of the correctness argument
for optimizations. In turn, OSR enables compiler transformations that
would otherwise be unsound. For instance, it allows the compiler to
speculate on \emph{likely} behaviors of the program, such as the dynamic type of
a variable. The speculation has to be guarded by a test and then the compiler can
rely on OSR, in case the speculation fails, to go back to the unoptimized
function. In other words, the
unlikely cases can be completely ignored by the optimizer and relegated to this
generic fall-back.

\paragraph{Difficulties}

OSR has been described as black magic due to the
non-conventional control-flow that it introduces. A significant part
of the complexity comes from the fact that most implementations do
not provide clean abstractions for OSR.  For example, extracting and
rewriting the program state, \ie steps (a) and (b) in \autoref{fig:osr}, are
often not separated cleanly.
Both of these two steps provide challenges, but for different
reasons. Extracting the program state is challenging due to low-level concerns.
We need very fine grained access to the internal state of the
computation at the OSR points. This access has to be provided by the
backend of our compiler, \eg by exposing how the execution state is mapped to
the hardware or the interpreter. On the other hand, mapping the extracted program 
state to a target state relies on the optimization providing the required
information.

From the optimizer's point of view, the challenge presented by OSR is
that the mapping information must be preserved during transformations.
Every transformation that is applied has to amass enough
meta-data for the state mapping to be well defined at every OSR point.
One approach to keeping the mapping valid is to represent it by
metadata or pseudo instructions inside the instruction
stream~\citep{dub13}.  For instance, the compiler used in this
work inserts so-called \c{Framestate} instructions in an early
translation phase.  These instructions capture the values on the
operand stack, local variables, and the program counter. They are the
description of the execution state needed for the mapping. 
While optimizing, the compiler keeps the frame states updated. Another
instruction, \c{Checkpoint}, acts as an anchor for the frame
states, and describes potential OSR exit points.  The compiler emits
them after each effect to allow for exits at most locations in the instruction
stream.

\added[comment={rev. (3) background}]{
An important challenge for speculation next to correctness is the hard to predict
performance. While it is folklore, few published works
investigate the resulting instabilities and their mitigation.
\citet{bar17} found a surprising amount of unexpected behavior, such as
performance degrading over time, or not stabilizing at all, in production VMs. These
results hint at the difficult trade-off of deciding when to optimize. Late
optimizations suffer a slow warm up, or the program even finishes before the optimizer
kicks in; eager optimizations risk mis-speculation. For instance \citet{url-v8-slow}
notes that deoptimization from mis-speculation in V8 can hurt the performance
especially early during page load.
\citet{zhe17} found that Graal sometimes should keep optimized code despite
deoptimization events. In other words, optimized code that is correct most of
the time can be faster than more generic code that is always correct.
}

\paragraph{Simplified OSR-in}

Whereas OSR-out relies on the ability to extract the source
execution state at many locations, OSR-in is simpler.
While one could arrange for OSR-in to enter optimized code in the
middle of a function, these entry points would limit optimizations and
would not be easy to implement if using an off-the-shelf code
generator such as LLVM (see \citet{lam13} for such an approach). Instead, one can compile a
continuation starting from the current program location to the end of
the current function.  This continuation is executed once and on the
next invocation the function is compiled a second time from the
beginning of the function. This approach simplifies the mapping of
execution states, as there is only one concrete state that needs to be
mapped instead of multiple abstract states at every potential entry
point. The current state is simply passed as an argument to the
continuation. This is a popular implementation choice following \citet{fin03}.

\paragraph{Implementation Choices for OSR-out}

The lowest overhead to peak performance for OSR exit points is achieved by
extracting the
execution state by an external mechanism. Typically at a defined location
execution is conditionally stopped and control transferred to an OSR-out implementation, \eg
by tail-calling it. The OSR-out implementation then uses the compiler's meta-data
to extract the run-time state from the registers and the run-time stack. A
simpler alternative implementation is to pass all the required state as
arguments to the OSR-out function. This approach generates more code,
as the state extraction is effectively embedded into the native code.
\added[comment={rev. (2) other lang, VMs}]{It would
be interesting to investigate the performance and memory-overhead trade-off for
using specialized code instead of meta-data for
deoptimization~\citep{dub14}.}

\subsection{A Short and Partial History of OSR}

OSR for deoptimization was pioneered in SELF by \citet{hol92}. At
first, the idea was simply to deoptimize code to provide a source-level
debugging experience. In that sense, it was a speculative optimization
on the assumption that debugging is not used.  Soon the idea was
applied to speculatively optimize for all kinds of assumptions, from
the stability of class hierarchies~\citep{pal01} to unlikely behavior
in general~\citep{bur99}, and providing more and more flexibility to
the optimizer in the presence of deoptimization~\citep{som06}.  We are
reaching the point where deoptimization is an off-the-shelf technique
\citep{lam13, mad14} that more and more compilers are relying on for
diverse purposes \citep{oda05, sch12, dub14, sta16, ari17, qun18}. The
common idea is that deoptimization leads the control-flow back to less
optimized code.  Deoptless provides an alternative option where we
split at deoptimization points and the specialization is instead 
increased.

OSR-in was first described by \citet{hol94} in their recompilation
strategy.  When a very small function is invoked often, they rather
recompile the caller and replace it using OSR-in. SELF, being an
interactive system, was concerned with compilation pauses.  Especially
given splitting-based optimizations that could lead to an explosion of
code size. \citet{cha91} address this issue by identifying uncommon
source-level control-flows and deferring their compilation.  \citet{sug03}
describe the natural extension of this idea where the deferred
compilation is implemented by means of OSR.  The Jikes RVM extensively
relied on OSR-in for profile-driven deferred compilation as described
by \citet{fin03}.  Deferred compilation can be understood as a
speculative optimization that assumes an unlikely source-level branch
is not taken.

The combination of OSR-out and OSR-in was explored by \citet{wim17}
to have an optimizing compiler act as the baseline compiler.  The
difference with our approach is that an OSR-out still ends in a
less optimized version of the code. To the best of our knowledge,
no other work employs polymorphic OSR-out.

\section{Deoptless}\label{sec:deoptless}

\emph{Deoptless} is a compilation strategy that explores the idea
of having a polymorphic OSR-out as a backup for failed speculation,
\begin{figure}
\begin{lstlisting}[label={lst:vector-sum},caption={Summing vectors}, style={R}]
 sum <- function() {
   total <- 0
   for (i in 1:length) total <- total + data[[i]]
   total
 }
\end{lstlisting}

  \bigskip

  \includegraphics[width=0.8\columnwidth]{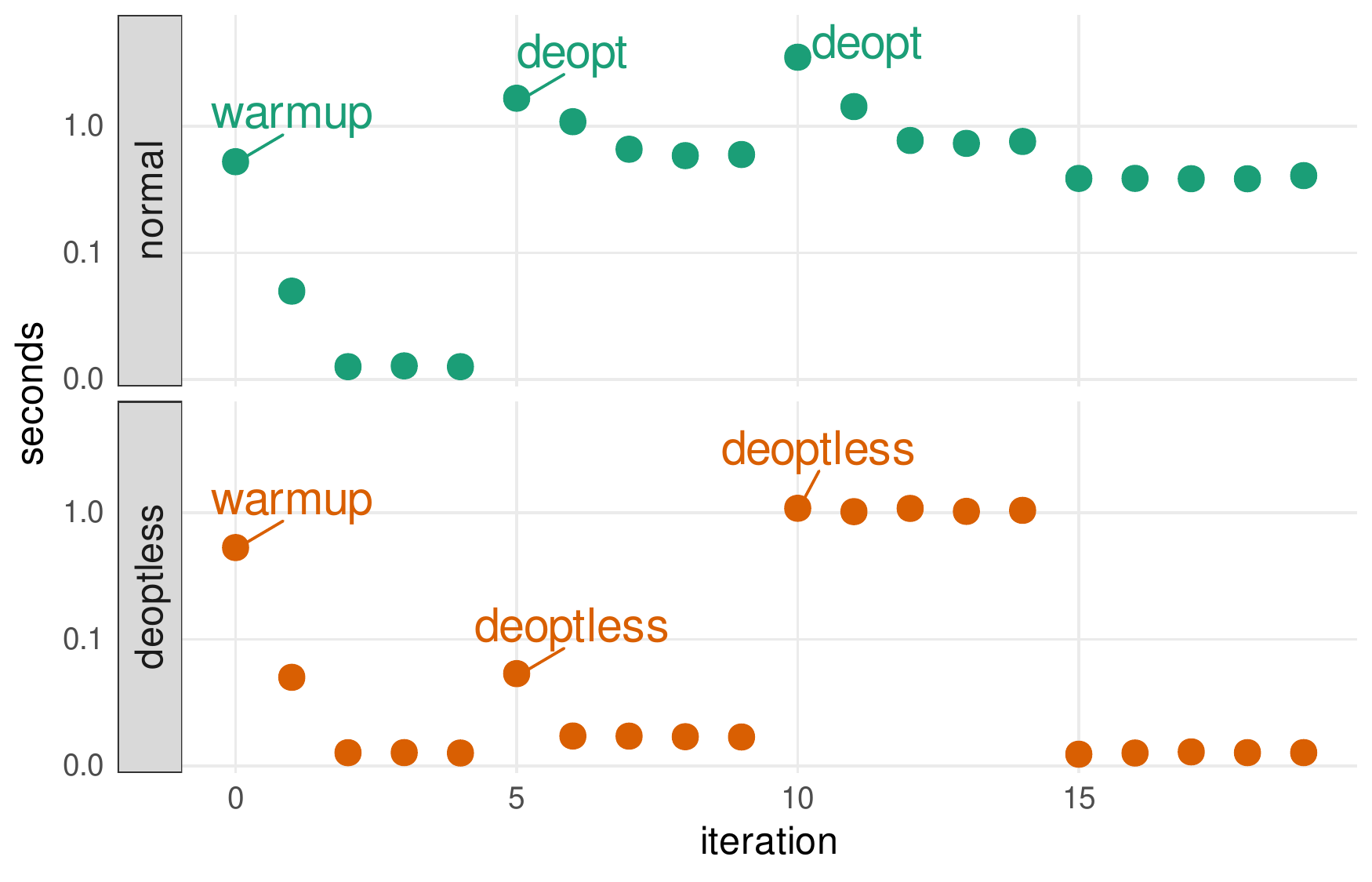}
  \caption{Performance comparison (log scale)}
  \Description[Performance example for deoptless]{Normal execution gradually
  slows down with every deoptimization event, deoptless retains performance.}
  \label{fig:comparison}

  \vskip -0.2em
\end{figure}
while retaining the version of the function that triggered deoptimization.
Consider the \c{sum} function of \autoref{lst:vector-sum} that
naively adds up all the elements in the \c{data} vector. Assume the function
is called in situations where the values of data change from float to
integer to complex numbers and back to float.
As a preview, we run this code in our implementation.
\autoref{fig:comparison} shows both normal executions and executions with deoptless.
We see the warmup time spent in the interpreter and compilation to faster native
code in the first phase with 5 iterations. Each of the following 3 phases (also
with 5 iterations each) correspond to a different type of \c{data} vector.
In the normal environment each change of the dynamic type results in
deoptimization, followed by slower execution. In deoptless, there is a slowdown
in the first iteration, as the continuation must be compiled, then code is fast again. 
Complex numbers are slow in both versions as their behavior is more involved. 
Finally, when the function
deals with floats again, deoptless is as fast as the first time, whereas the original version
is stuck with slow code. We show this example here to motivate the technique
and give an intuition for our goals and the expected gains. This graph
effectively illustrates many of the trade-offs with deoptless that we are aware of,
and we'll discuss it again in detail at the end of the section.

\subsection{Approach}

\begin{figure}[t]
  \includegraphics[width=0.95\columnwidth]{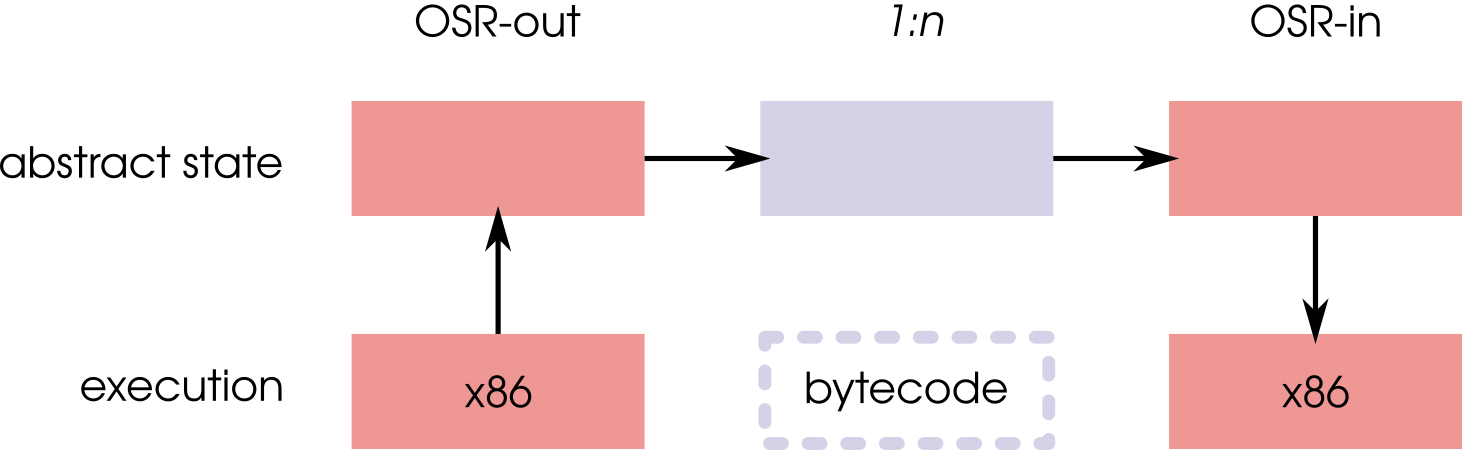}
  \caption{Deoptless combines OSR-out with OSR-in}
  \Description{Abstract control-flow of a deoptless OSR.}
  \label{fig:osr_deoptless}
\end{figure}

Conceptually, deoptless performs OSR-out and OSR-in in one step, to achieve
optimized-to-optimized and native-to-native handling of failing speculation.
As can be seen in \autoref{fig:osr_deoptless} this is realized by following an OSR-out
immediately with an OSR-in.
By performing this transition directly, it is possible to never leave optimized code.
For deoptless, the OSR-in must be implemented by compiling an optimized
continuation, specifically for that particular OSR exit point. The key idea is
that we can compile multiple specialized continuations, depending on the failing
speculation --- and in general, depending on the current state of the
execution. The continuations are placed in a dispatch table to be reused in future
deoptimizations with compatible execution states.

We effectively turn deoptimization points into assumption-polymorphic
dispatch sites for optimized continuations.
If the same deoptimization
exit point is taken for different reasons, then, depending on the reason,
differently specialized continuations are invoked.
Going back to \autoref{lst:vector-sum}, the failing
assumption is a typecheck. Given earlier runs, the compiler speculates that
\c{data} is a vector of floats. This assumption allows us to access the vector
as efficiently as a native array. Additionally, based on that assumption, the
\c{total} variable is inferred to be a float scalar value and can be unboxed.
When the variable becomes an integer, this speculation fails. Normally we would
deoptimize the function and continue in the most generic version, \eg in the baseline
interpreter. Deoptless allows us to split out an
alternate universe where we speculate differently and jump to a continuation optimized for that case.

\paragraph{Dispatching}

We keep all deoptless continuations of a function in a common dispatch table. At
a minimum the continuation we want to invoke has to be compiled for the same
target program location. But we can go further and use the current program
state, that we extracted from the origin function for OSR, to have many
specialized continuations for the same exit. In order to deduce which continuations are compatible
with the current program state we employ a context dispatching mechanism. In this framework, code is optimized under a context of
assumptions. Such an optimization context $C$ is a predicate over the program state with an
efficiently computable partial order
$C_1 < C_2~\mathit{iff}C_1\Rightarrow C_2$.
A context is called current with respect to a state $S$ if $C(S)$ holds.
To choose a continuation, we take the current state at the OSR exit point, we
compute a current context $C$ for it, and then select a continuation compiled
for a context $C'$, such that $C < C'$. If there is no such continuation
available, or we find the available ones to be too generic given the current
context, we can choose to compile a new continuation and add it to the dispatch
table.

In our implementation we add an abstract description of the deoptimization reason,
such as "typecheck failed, actual type was an integer vector", to the context.
Our source states are expressed in terms of the state of the bytecode interpreter.
Therefore, the deoptimization context
additionally contains the program counter of the deoptimization point, the names and types of
local variables, and the types of the variables on the bytecode stack.
As mentioned, contexts are partially ordered. Our contexts are only comparable if they have the same deoptimization target, the
same names of local variables, the same number of values on
the operand stack, and a compatible deoptimization reason.
This means, for instance, that a deoptimization on a failing
typecheck is not comparable with a deoptimization on a failing dynamic
inlining, and thus we can't reuse the respective continuation.
Or, if there is an additional local variable that does not exist in the continuation context.
Comparable contexts are then sorted by the degree of
specialization. For instance, they are ordered by the subtype relation of the
types of variables and operands. If the continuation is compiled
for a state where \c{sum} is a number, then it can
for example be called when the variable \c{sum} holds an integer or a
floating-point number. Or, if we have a continuation for a typecheck, where we
observed a float vector instead of some other type, then this continuation
will be compatible when we observe a scalar float instead, as in R
scalars are just vectors of length one.

Dispatching is based on the execution states of the source code of the optimizer,
\eg in our case states of a bytecode interpreter. This does not mean that deoptless
requires these states to be materialized. For instance when dispatching on the type of
values that would be on the operand stack of the interpreter at the
deoptimization point, they are not actually pushed on the stack. Instead their
type is tested where they currently are in the native state.

\subsection{Discussion}

Deoptless does not add much additional
complexity over OSR-out and OSR-in to an implementation. There are some
considerations that will be discussed when we present our prototype in the next
section. Most prominently,
OSR-out needs to be more efficient than when it is used only for deoptimization,
because we expect to trigger OSR more frequently when dispatching to optimized
continuations.
Currently our proof-of-concept implementation is limited to handle
deoptimizations where the origin and target have one stack frame, \ie we do not
handle inlined functions. This is not a limitation of the technique, but rather
follows from the fact that also the OSR-in implementation currently has the same
limitation. We can therefore not answer how well deoptless would perform in the
case of inlined functions.

There are also a number of particular trade-offs, which are
already visible in the simple example in \autoref{fig:comparison}.
Going through
the four phases of the example, we can observe the
following. In the first phase both implementations warm up equally fast. There
is no difference, as there is also no deoptimization event up to this point. In the second phase,
when the type changes to float, the normal implementation triggers a
deoptimization, we fall back to the interpreter and it takes some time for the
code to be recompiled. This replacement code is more generic as it can handle
floats and integers at the same time and it is much slower than the float-only
case. The effect is inflated here due to the fact that our particular compiler
supports unboxing only if the types are static. This can be seen in the first
phase of the deoptless variant, where a specialized continuation for floats is
compiled and executed very efficiently. We see a small overhead over the integer
case, that is due to the dispatch overhead of deoptless.
Next, in the third phase, yet another specialized
continuation is compiled, this time for the \c{data} vector being a generic R
object. While we avoid going back to the interpreter yet again, this continuation is
slower at peak performance than the generic version from the normal execution.
This is not a fundamental limitation, but does exemplify a difficulty with
deoptless that we will get back to: deoptless operates on partial
type-feedback from the lower tier. Because the remainder of the \c{sum} function has never been
executed with the new type, we cannot fully trust the type-feedback when
compiling
the continuation, as it is likely stale to some extent. We address the problem with a selective
type-feedback cleanup and inference pass, which can, as in this case, lead to less optimal
code. In the final phase of the benchmark deoptless greatly outperforms the
normal implementation. That is because in deoptless we are running the same code
again as in the first phase, as this code was never discarded. On the other hand
in the normal case we replaced the sum function in-place and it is now much more
generic and slow.

\section{Implementing Deoptless in the \R JIT}\label{sec:rsh-deoptless}

In this section we detail our efforts to introduce deoptless to \R, an
optimizing just-in-time compiler for the R language. \R
features a two-and-half-tier optimization strategy, with the two traditional
tiers, a bytecode interpreter and a native optimizing compiler. Additionally, 
it falls back to the AST interpreter from GNU R, as R is
a language with a fairly large number of features, some of which are not yet supported.

\R has an existing OSR-out
implementation to transition from native code to the interpreter in case of
mis-speculation. \R uses assumptions about
the stability of call targets, the declared local variables of closures,
uncommon branches, primitive types, and loops over integer sequences
to speculatively optimize code.
Speculation is an inherent feature of its intermediate representation (IR).
It is expressed by \c{Assume} instructions which
function similarly to \c{asserts}, the difference being that failing assumptions
are silently handled by deoptimization. The conditions guarded by \c{Assume} instructions
are used by various optimizer passes.

\R also features a recent OSR-in implementation, a direct side-product of our work to
implement deoptless. It can be used to tier-up from the interpreter to optimized code,
and is triggered in long-running loops. Supporting
OSR-in adds little complexity to the compiler. The relevant patch adds 300 and changes
600 lines of code. Mainly, the bytecode to IR translation has to support starting at an
offset, and the current values on the interpreter's operand stack need to be passed into the
optimized continuation.

Deoptless can be easily implemented on top of an existing implementation
of OSR-out and -in. The patch adds 600 and changes 300 lines of code. Compilation
is straightforward using the OSR-in implementation. The additional complexity
stems from defining optimization contexts suited for deoptless, and then
context dispatching over these contexts. Finally, a new type-feedback inference
and cleanup pass is required. In the original case, 
the interpreter collects new run-time feedback after a deoptimization and before code is reoptimized. 
With deoptless, we try to recompile right after a failing assumption, not having a chance to
capture later, secondary changes to the program state that we need to update our assumptions about the code.
The feedback inference pass tries to remove profile data which is likely
invalid after the failing assumption, and infer new feedback from the remaining data.
Before describing our deoptless implementation in detail, we
present the implementation of OSR-out and -in.

\subsection{OSR-out}

In the intermediate representation of the compiler, OSR exit points are represented by
\c{Checkpoints}. A checkpoint instruction is the anchor that keeps OSR origin and
target code in sync. Each one belongs to a description of the target execution state,
represented by a \c{Framestate} instruction.

Of particular importance is the \c{Assume} instruction, which represents a
run-time assumption made by the compiler to be used for optimizations. An
\c{Assume} refers to a \c{Checkpoint} that can be used if its guard fails.
An example can be seen in \autoref{lst:pir-osr-exit}.
In R, the local variable scope is a first-class object called the
\emph{environment} --- we refer to \citet{dls19} for a detailed
explanation --- and it has to be materialized on deoptimization.
The deferred instructions at label \c{D1} represent the materialization of
the environment and describe the \c{Framestate} required to exit
from this \c{Checkpoint}. The frame describes an \R bytecode
execution context at program counter location \c{15} with an R environment where
\c{sum} is bound to \c{0}.
In general, \c{Framestate} instructions can be chained to describe the states of multiple
inlined functions. The basic block \c{D1} can contain arbitrary deferred instructions 
that are executed only upon deoptimization. This is frequently used to defer
computations which are not needed in the optimized code. As an
example, R has a call-by-need semantics, and function arguments are passed as thunks,
so-called promises. After inlining, the actual structure to hold the delayed
computation typically has to be created only on deoptimization, and the
instructions for creating promises can be delayed into deoptimization branches.

\begin{figure}\centering\begin{lstlisting}[style=pir,label={lst:pir-osr-exit},caption={OSR exit point in \R}]
   %c = Checkpoint D1
        ...
   %d = LdVar(data, GlobalEnv)
   %t = IsType[real](%d)
        Assume(%t, %c)
        ...
 D1:
   %e = MkEnv(sum=0)
   %f = Framestate(%e, pc=15)
        Deopt(%f)
\end{lstlisting}\vskip -0.8em\end{figure}

In \R, OSR exits are not performed by externally rewriting stack frames. Instead, an OSR exit point is
realized as a function call. Let us consider the OSR exit point in
\autoref{lst:pir-osr-exit}. The backend of the \R compiler generates code using LLVM.
As can be seen in \autoref{lst:llvm-osr-exit}, the \c{Assume} is lowered to a conditional branch 
and the OSR exit is lowered to a tail-call.
\begin{figure}\begin{lstlisting}[style=llvm,label={lst:llvm-osr-exit},caption={OSR exit from \autoref{lst:pir-osr-exit} in LLVM}]
         br %isType, cont, osr
  osr:
         ...
    %f = alloca FrameState
    %r = alloca Reason
       ; store current function,
       ; frame contents, and more
       ; metadata into %f and %r
    %a = tail call void @deopt(%f, %r)
         ret %a
  cont:
\end{lstlisting}\vskip -0.8em\end{figure}
\noindent The \c{osr} block executes all the deferred instructions,
notably it populates buffers for the local variables captured by
the framestate and the deoptimization reason.
Finally, a \c{deopt} function is called. This primitive performs the
actual deoptimization, \ie it
invokes the interpreter, or, in the case of deoptless, dispatches to an
optimized continuation.

The \c{deopt} primitive is able to recreate multiple interpreter contexts as we
can see in the pseudocode in \autoref{lst:deopt-impl}.
\begin{figure}\begin{lstlisting}[style=C,label={lst:deopt-impl},caption={Pseudocode for deoptimization implementation}]
  Value deopt(FrameState* fs, Reason* r) {
    logDeoptimization(r);
    pushInterpreterState(fs);
    if (fs->next)
      push(deopt(fs->next, r));
    return interpret(fs->pc, fs->env);
  }
\end{lstlisting}\vskip -0.8em\end{figure}
First, the outer interpreter context is synthesized, \ie the necessary values 
pushed to the interpreter's operand stack. Then, the inner frames are recursively 
evaluated, their results also pushed to the operand stack, as expected by the outer
frame. Finally, the outermost code is executed, and the result returned to the
deoptimized native code, which directly returns it to its caller.

The \c{osr} basic block in \autoref{lst:pir-osr-exit}, as well as the \c{deopt}
call, are marked
\emph{cold} in LLVM. This should cause LLVM optimization passes and code
generation to layout the function in such a way that the impact of the \c{osr}
code on the performance of the rest of the function is minimal.
However, the mere presence of the additional branch might interfere with LLVM
optimizations, and other OSR implementers therefore chose to use the LLVM \c{statepoint} primitive.
The statepoint API provides access to metadata describing the stack layout of the
generated function. This stack layout allows an external deoptimization
mechanism to read out the local state without explicitly capturing it in LLVM code.
This is a trade-off, and the impact is in our opinion limited. For example, in
the concrete case of \autoref{lst:vector-sum}, we were
not able to measure any effect on peak performance. In fact, when we unsoundly dropped all deoptimization
exit points in the backend, the performance was unchanged. There was, however, an
effect on code size with an overhead of 30\% more LLVM instructions.
The implementation strategy of using explicit calls to \c{deopt} for \R was chosen for ease of implementation long
before deoptless was added. In a lucky coincidence, this strategy is very
efficient in extracting the internal state of optimized code compared to an
external deoptimization mechanism, and therefore very well suited for
deoptless.

\subsection{OSR-in}

Together with deoptless, a new OSR-in mechanism was introduced in \R,
since the codebase can be mostly shared. OSR-in allows for a transition from
long-running loops in the bytecode interpreter to native code. To that end, a
special continuation function is compiled, starting from 
the current bytecode, which is used only once for the OSR-in. The
full function is compiled again from the beginning the next time it is called.
This avoids the overhead of introducing multiple entry-points
into optimized code, for the price of compiling these functions twice. Since
OSR-in is not a very frequent event, the trade-off is reasonable. 

The mechanism is triggered by counting the number of backward jumps in the interpreter.
When a certain number of loop iterations is reached, the remainder of the function is 
compiled using the same compiler infrastructure that is used to compile whole functions.
The only difference is that we choose the current program counter value as an entry point for
the translation from bytecode to IR. Additionally, we pre-seed the abstract stack used by the frontend of
the \R compiler with all values on the interpreter's operand stack.
In other words, the resulting native code will receive the current contents of 
the operand stack as call arguments. OSR adds the
lines, shown in \autoref{lst:osr-in-impl}, to the implementation of the branch bytecode.

\begin{figure}
\begin{lstlisting}[style=C,label={lst:osr-in-impl},caption={Pseudocode for OSR-in implementation}]
 case Opcode::branch: {
   auto offset = readImmediate();
   if (offset < 0 && OSRCondition()) {
     if (auto fun = OSRCompile(pc, ...)) {
       auto res = fun(...);
       clearStack();
       return res;
     }
   }
   ...
 }
\end{lstlisting}\vskip -0.8em\end{figure}

An interesting anecdote from adding OSR-in to \R is that out of all the
optimization passes of the normal optimizer, only dead-store elimination was
unsound for OSR-in continuations. The reason is that objects can already
escape before the OSR continuation begins, and thus escape analysis would
mistakenly mark them as local.

\subsection{Deoptless}

Our implementation underscores the point that adding deoptless to a VM with an existing 
implementation of OSR-in and OSR-out requires only minimal changes. Starting with
the code in \autoref{lst:deopt-impl}, we extend it as shown in \autoref{lst:deoptless-impl}.
In this listing we see five functions that we'll detail to explain the
implementation. \c{deoptlessCondition} decides if deoptless should be
attempted. Certain kinds of deoptimizations do not make sense to be handled, and also
our proof of concept implementation has limitations and is not able to handle
all deoptimizations. Then, \c{computeCtx} computes the current optimization
context and \c{dispatch} tries to find an existing continuation that is
compatible with the current context. \c{recompile} is our recompilation
heuristic that decides if a continuation, while matching, is not good enough.
Next, the \c{deoptlessCompile} function invokes the compiler to compile a new deoptless
continuation. Finally, we call the compiled continuation, directly passing the
current state. The calling convention is slightly different from normal OSR-in.
As we are originating from native code the values can have native representations, 
whereas if we originate from the interpreter all values are boxed heap objects.

\begin{figure}\begin{lstlisting}[style=C,label={lst:deoptless-impl},caption={Pseudocode for deoptless implementation}]
 Value deopt(FrameState* fs, Reason* r) {
   if (deoptlessCondition(fs, r)) {
     auto ctx = computeCtx(fs, r);
     auto fun = fs->fun->deoptless->dispatch(ctx);
     if (!fun || recompile(fun, ctx))
       fun = deoptlessCompile(ctx);
     if (fun)
       return fun(fs);
   }
   // Rest same as normal deopt
 }
\end{lstlisting}\vskip -0.8em\end{figure}

\paragraph{Conditions and Limitations}

As mentioned, deoptless is not applied to all deoptimization events. First of
all, some deoptimizations are rather catastrophic for the compiler and prevent most
optimizations. An example would be an R environment (the dynamic
representation of variable scopes) that leaked and was non-locally modified. Under
these circumstances the \R optimizer cannot realistically optimize the code
anymore.
Second, when global assumptions change, \eg a library
function is redefined, we must assume that the original code is permanently
invalid and should actually be discarded.
Furthermore, we also prevent
recursive deoptless. If a deoptless continuation triggers a failing
speculation, then we give up and perform an actual deoptimization.
There are also some cases which are
not handled by our proof of concept implementation. The biggest limitation is
that we do not handle cases where more than one framestate exists, \ie we
exclude deoptimizations inside inlined code. This is not an inherent limitation,
and we might add it in the future, but so far we have avoided the implementation
complexity.

\paragraph{Context Dispatch}

Deoptless continuations are compiled under an optimization context, which
captures the conditions for which it is correct to invoke the continuation. The
context is shown in \autoref{lst:deopt-context} in full.
It contains the deoptimization target, the reason, the types of values on the
operand stack, and the types and names of bindings in the environment. The deoptimization
reason represents the kind of guard that failed, as well as an abstract
representation of the offending value. For instance, if a type guard failed, then
it contains the actual type, if a speculative inlining fails, it contains the
actual call target, and so on.

\begin{figure}\begin{lstlisting}[style=C,label={lst:deopt-context},caption={Deoptless optimization context}]
 struct DeoptContext {
   Opcode* pc;
   Reason reason;
   unsigned short stackSize;
   unsigned short envSize;
   Type stack[MAX_STACK];
   tuple<Name, Type> env[MAX_ENV];
   bool operator<= (DeoptContext& other);
 };
\end{lstlisting}\vskip -0.8em\end{figure}

The (de-)optimization context is used to compile
a continuation from native to native code, so why does it
contain the \c{Opcode* pc} field, referring to the bytecode instead? Let's reexamine 
\autoref{fig:osr_deoptless}. The state is extracted from native code and
directly translated into a target native state. However, logically, what connects
these two states is the related source state. For instance, the bytecode program
counter is used as an entry point for the \R compiler. The bytecode state is never
materialized, but it bridges the origin and target native states on both ends of deoptless.

Contexts are partially ordered by the \c{<=} relation. The relation is defined such that 
we can call a continuation
with a bigger context from a smaller current context. In other words, the
\c{dispatch} function from \autoref{lst:deoptless-impl} simply scans the
increasingly sorted dispatch table of continuations for the first one with a context \c{ctx'} such that
\c{ctx <= ctx'}, where \c{ctx} is the current context. The dispatch tables uses a
linearization of this partial order. The linearization currently does not favor a
particular context, should multiple optimal ones exist. For efficiency of the
comparison and dispatching, we limit the maximum number of elements on the stack
to 16 and environment sizes to 32
(states with bigger contexts are skipped), and only allow up to 5 continuations in the dispatch table.

\paragraph{Compilation and Calling Convention}

Compilation of deoptless continuations is performed by the normal \R optimizer
using the same basic facilities as are used for OSR-in. Additionally,
information from the \c{DeoptContext} is used to specialize the code further.
For instance, the types of values on the operand stack can be assumed stable by
the optimizer, since context dispatch ensures only compatible continuations are
invoked. The calling convention is such that the R environment does not
have to be materialized. The local R variables, which
are described by \c{Framestate} and \c{MkEnv} instructions at the
deoptimization exit point, are passed in a buffer struct.

\paragraph{Incomplete Profile Data}

An interesting issue we encountered is
incomplete type-feedback. As depicted in \autoref{fig:compare-deopt},
normally after a deoptimization event, the execution proceeds in the lower-tier, \eg in
the interpreter, which is also responsible for capturing
run-time profile data, such as type-feedback, branch frequencies, call targets,
and so on. When an assumption fails, this typically indicates that some of this
profile was incomplete or incorrect and more data is needed.
In deoptless we can't collect more data before recompiling, therefore we lack
the updated feedback. If we were to compile the
continuation with the stale feedback data, most
probably we would end up mis-speculating. For instance
if a typecheck of a particular variable fails, then the type-feedback for
operations involving that variable is probably wrong too. We
address this problem with an additional profile data cleanup and inference pass.

The cleanup consists of marking all feedback that is connected to the program
location of the deoptimization reason, or dependent on such a location, as
stale. Additionally we check all the feedback against the current run-time state
and mark all feedback that is contradicting the actual types.
Additionally, we insert available information from the
deoptimization context. For instance, if we deoptimize due to a typecheck,
then this step injects the actual type of the value that caused the guard to fail.
Finally we use an inference pass on the non-stale feedback to fill in the blanks.
For inference we reuse the static type
inference pass of \R, but run it on the type feedback instead
and use the result to update the expected type.
These heuristics work quite well for the evaluation in the next
section, however, it is possible that stale feedback is still present and causes
us to mis-speculate in the deoptless continuation, which leads to the function being
deoptimized for good.

\paragraph{Transferability}

\added[comment={rev. (2) other lang, VMs}]{
The description of deoptless focused on our current
implementation for concreteness. However, the technique generalizes and any
language implementation using speculative optimizations and deoptimization could
employ it. The only requirement is sufficiently efficient OSR-out and -in
support. To bridge the two, there needs to be some efficient way of converting
the extracted state of the OSR-out to match the calling convention of the OSR-in
fragment. For dispatching, many options are conceivable. We recommend to at least
specialize on the types of the variables captured by the deoptimization
metadata.
}

\section{Evaluation}\label{sec:eval}

Let us now turn to the question of how well deoptless works with respect to our
stated goals. Our aim is to
\begin{enumerate}
  \item reduce both the frequency and amplitude of the temporary slowdowns due
    to deoptimizations, and
  \item prevent the long-term over-generalization of code due to deoptimization and
    recompilation.
\end{enumerate}
Following these stated goals, we try to answer the following questions:
(1) Given the same deoptimization triggering events, what is the speedup of using deoptless?
(2) Is deoptless able to prevent over-generalization?

The nature of deoptless makes it challenging to answer these questions as the events we are
trying to alleviate are by definition rare. In particular the code produced by
\R is not going to cause many deoptimizations in known benchmark suites.
Therefore, we decided to perform our main evaluation of deoptless on the worst-case situation, where we
randomly fail speculations.
Secondly, we will
evaluate deoptless on bigger programs, with known deoptimizations, due to the
nature of their computations.

\paragraph{Methodology}

Experiments are run on a dedicated benchmark machine, with all background
tasks disabled. The system features an Intel i7-6700K CPU, stepping 3, microcode
0xea with 4 cores and 8 threads, 32 GB of RAM and Ubuntu
18.04 on a 4.15.0-151 Linux kernel. Experiments are built as
Ubuntu 20.04.1 based containers, and executed on the Docker runtime 20.10.7,
build f0df350.\footnote{We use a containerized environment to automate
measurements and verified that it does not distort the results.} Measurements are recorded repeatedly and we keep a historical
record to spot unstable behavior.
For some of the experiments we use the major benchmarks from the \R
benchmark suite~\citep{oopsla20c}.

\subsection{Speedup over Deoptimization}

\begin{figure}[t]
  \includegraphics[width=1.0\columnwidth]{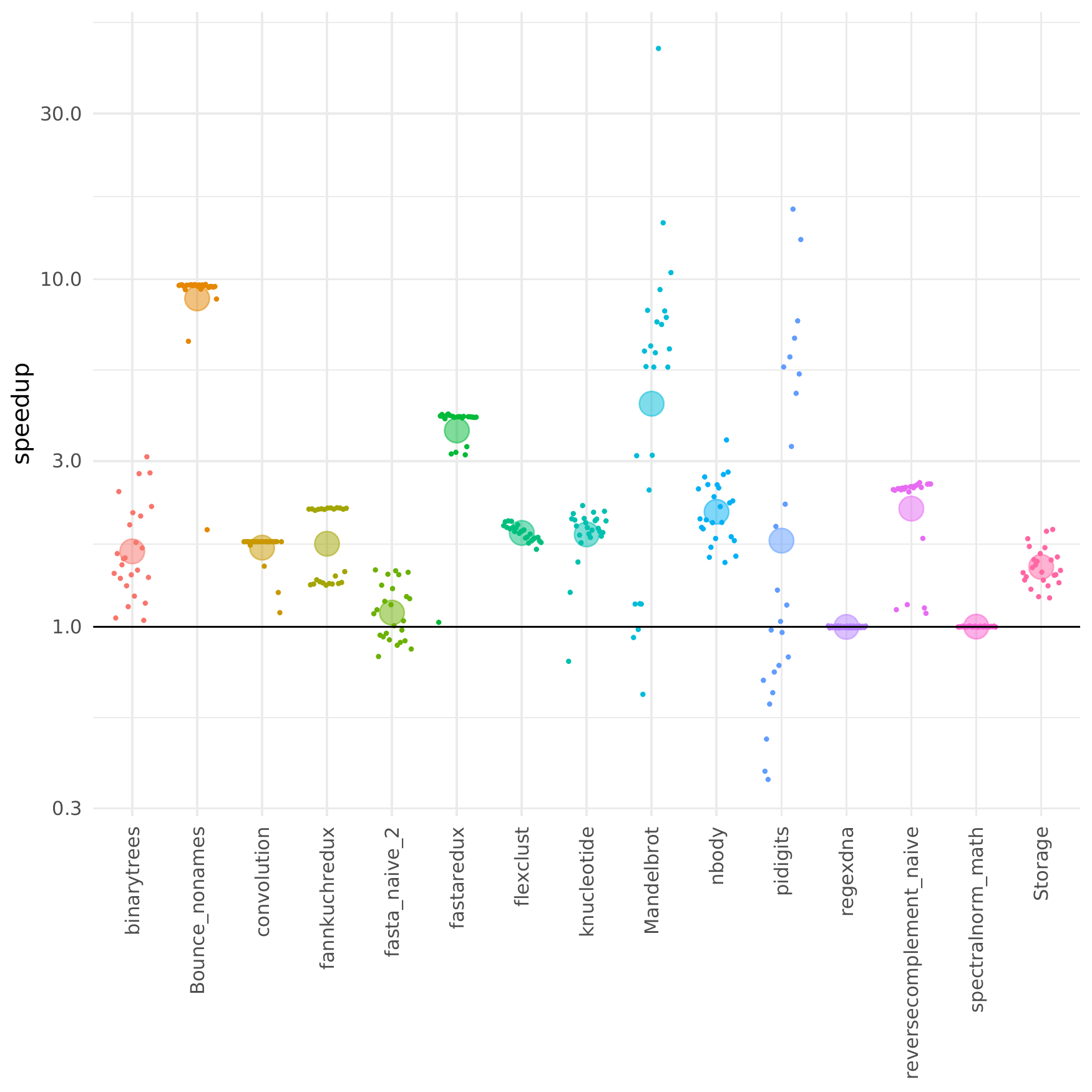}
  \caption{Deoptless speedup on mis-speculation (log scale)}
  \Description{Graph of the average speedup and each individual measurement.}
  \label{fig:deoptless-speedup}
  \medskip
\end{figure}

First we want to evaluate the performance gains of deoptless from avoiding
deoptimization alone. To that end we take the default \R main benchmark suite and
randomly invalidate 1 out of 10k assumptions. To be precise, we only trigger
deoptimization without actually affecting the guarded fact. This is achieved by
instrumenting the compiler to add a random trigger to every run-time check of an
assumption. This is an already existing feature of \R used in development to test the
deoptimization implementation. Enabling this mode causes a large slowdown of
the whole benchmark suite. We then measure how much of
that slowdown can be recovered with deoptless. Note that this is a worst-case
scenario that does not evaluate
the additional specialization provided by deoptless, as the triggered
deoptimizations largely correspond to assumptions that in fact still hold.
We run this experiment with 30 in-process iterations times 3 executions.
The results are presented in \autoref{fig:deoptless-speedup}. The
large dots in the graph show the speedup of deoptless over the baseline on a log
scale on average.
Improvements range from $\perfMin\times$ to
\replaced[comment={convolution had unstable baseline}]{$\perfMax\times$}{$50\times$},
with most benchmarks
gaining by more than \replaced[comment={bugfixes}]{$\perfMed\times$}{$1.6\times$}. The small dots represent in-process iterations from left
to right, averaged over all executions. We exclude the first 5 warmup
iterations, as they add more noise and only slightly affect the averages. Normalization is done for every dot
individually against the same iteration number without deoptless. From the main benchmark suite we
had to exclude the \c{nbody_naive} benchmark, as it takes over one
hour to run in the deoptimization triggering test mode. Though, we would like to
add, that with deoptless this time is cut down to less than five minutes.
Overall this experiment shows that deoptless is significantly faster then
falling back to the interpreter for the \R benchmark suite.

\paragraph{Memory Usage}

\added[comment={rev. (1) memory usage}]{
Deoptless causes more code to be compiled, which can lead
to more memory being used. The R language
is memory hungry
due to its value semantic,
and running more optimized code leads to fewer allocations.
Thus we expect deoptless to not use more memory overall. In this
worst-case experiment with randomly failing assumptions we measured a median
decrease of \memMed\% in the maximum resident set size. There is one outlier increase
in flexclust by \memMax\% and several decreases, the largest being \memMin\% in
fannkuchredux. The trade-off could be different for other languages or implementations.
However, the overhead can always be limited by the maximum number of deoptless
continuations.
}

\subsection{Benchmarks}

In the following we investigate the effects of deoptless on a selection of
benchmarks with known deoptimization events.

\paragraph{Volcano}

\begin{figure}[t]
  \includegraphics[width=0.7\columnwidth]{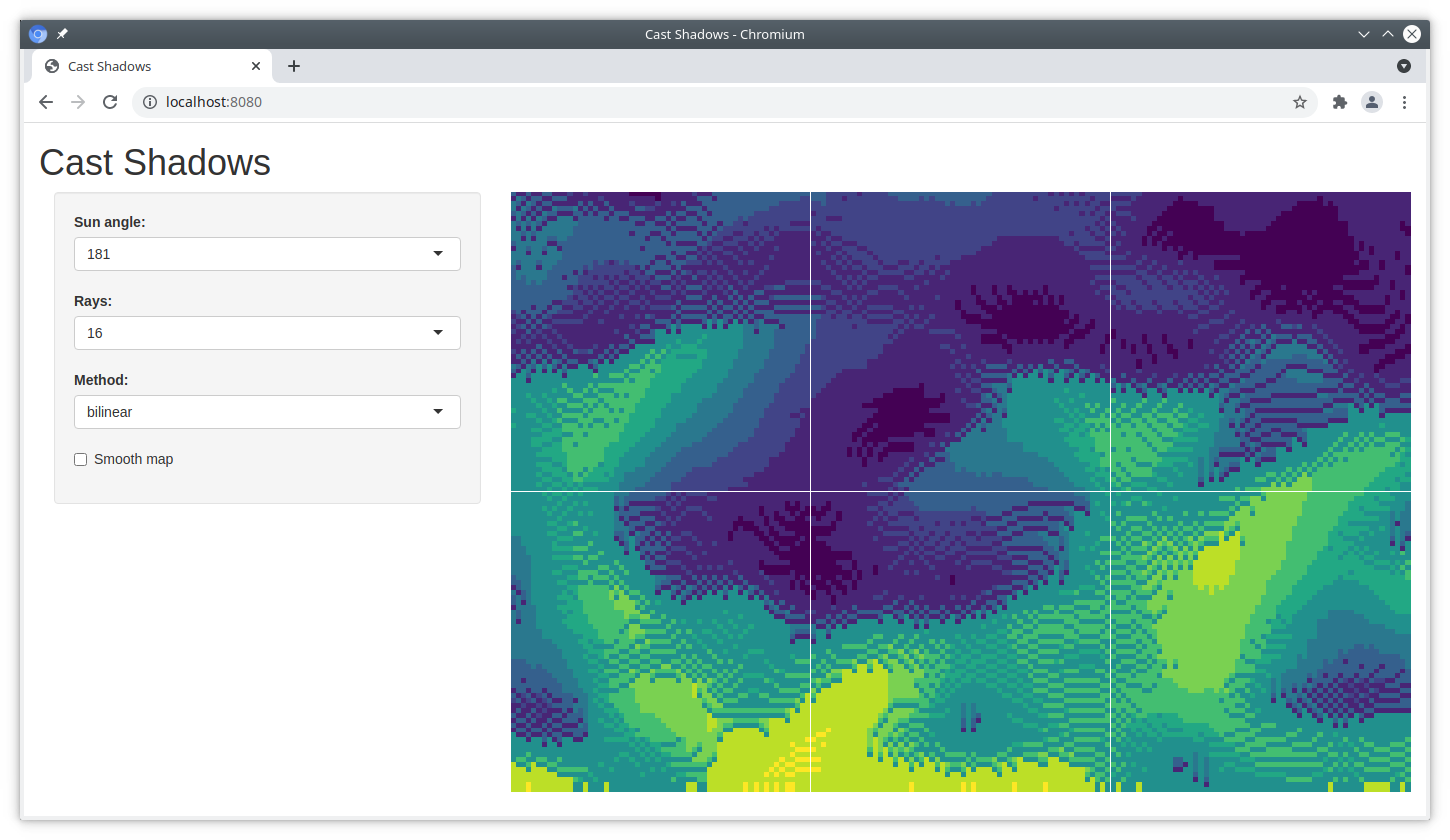}
  \caption{Volcano rendering shiny app}
  \Description{A screenshot of the application.}
  \label{fig:volcano-app}
  \bigskip
  \ifpreprint\bigskip\else\fi
  \includegraphics[width=0.7\columnwidth]{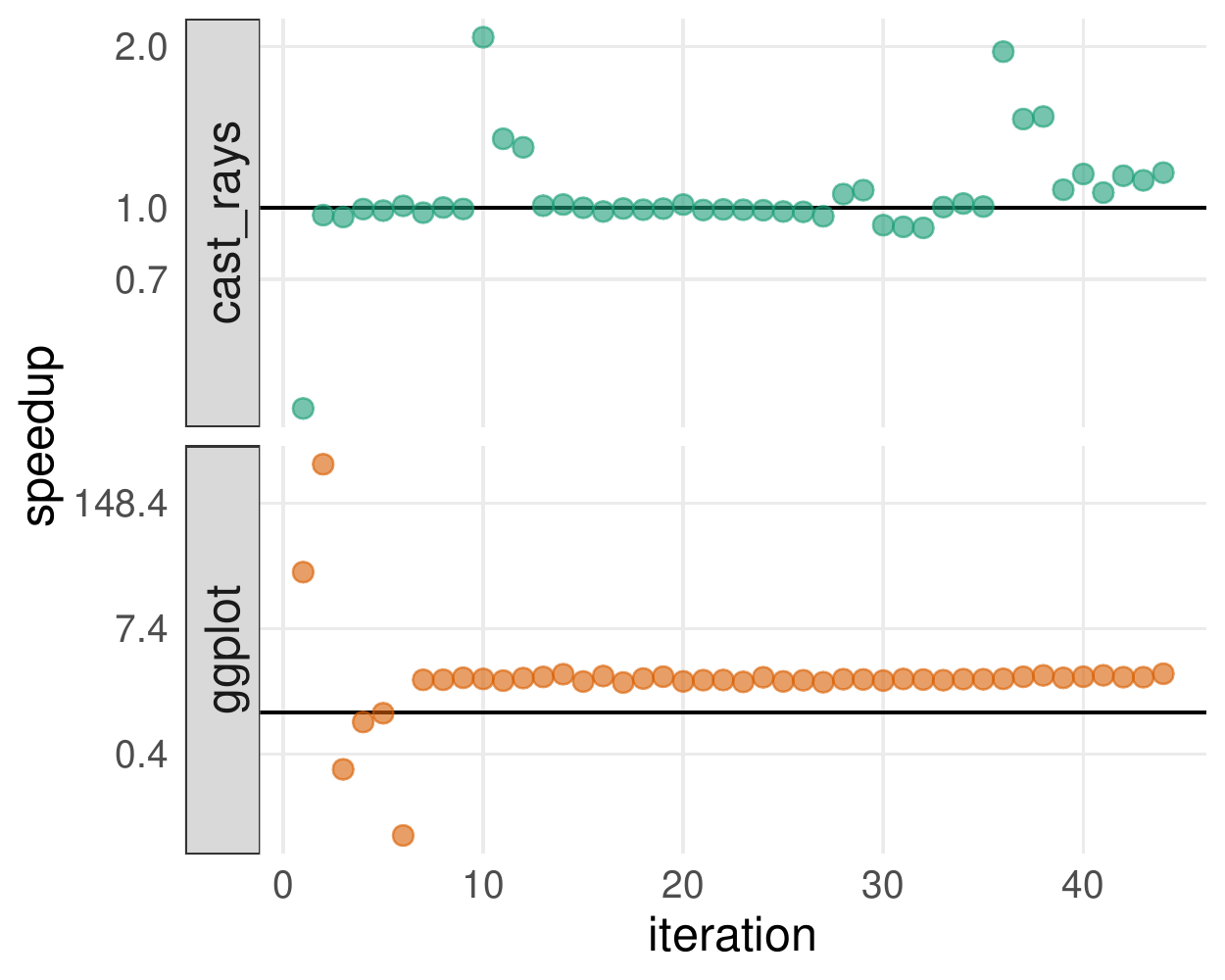}
  \caption{Volcano app speedup (log scale)}
  \Description{Graph that shows how the interactive session has less severe
  slowdowns with deoptless.}
  \label{fig:volcano-comparison}
  \bigskip
  \ifpreprint\bigskip\else\fi
  \includegraphics[width=0.7\columnwidth]{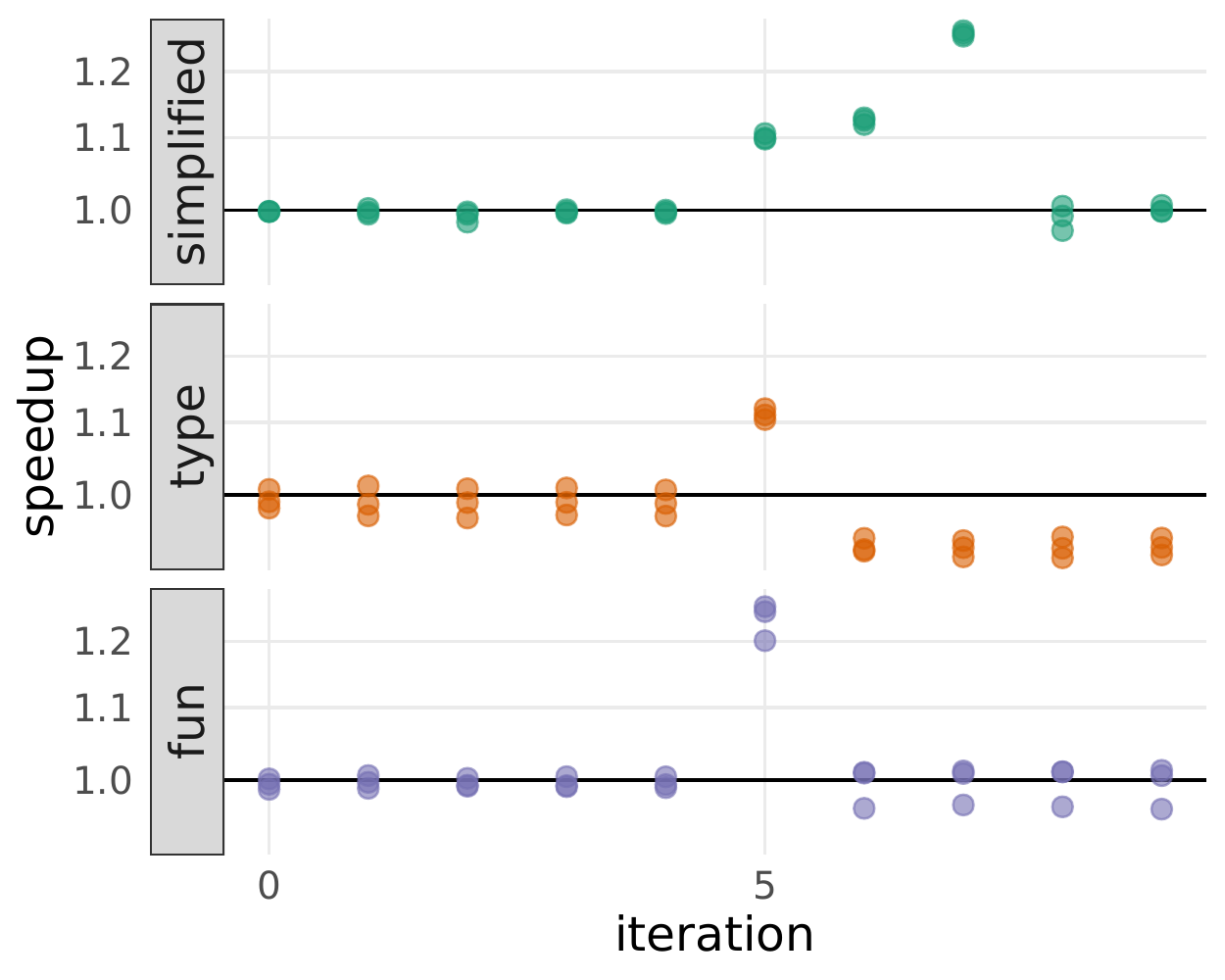}
  \caption{Ray-tracings with deoptimization at iteration 5}
  \Description{Graph that shows how the individual parts of the volcano
  application have less severe slowdowns with deoptless.}
  \label{fig:volcano-comparison-2}
\end{figure}

Deoptimizations can happen when user interaction leads to events which are not
predictable.
To demonstrate the effect we package a ray-tracing
implementation~\citep{url-throwing-shade} as a shiny
app~\citep{package-shiny} shown in
\autoref{fig:volcano-app}. It allows the user to
select properties, such as the sun's position, selecting the functions for
numerical computations and so on. The app renders a picture using
ggplot2~\citep{package-ggplot2} and the aforementioned
ray-tracer with a height-map of a volcano.
At the core of the computation is a loop nest which incrementally updates the
pixels in the image, by computing the angle at which rays intersect the terrain.
We record two identical sessions of a user clicking
on different features in the app. We then measure for each interaction how long the
application takes to compute and render the picture. In \autoref{fig:volcano-comparison} we
show the relative speedup of deoptless for that interactive session, separate
for the ray-tracing and the rendering step. The application exhibits
deoptimization events when the user chooses a different
numerical interpolation function. Deoptless results
in up to $2\times$ faster computations for these particular iterations. In
general deoptless always computes faster, except for one warmup iteration with a
longer compile pause. The produced image is then rendered by ggplot where we see
deoptless' ability to prevent over-generalization. The
code consistently runs about $2.5\times$ faster after warmup.

We also investigate other kinds of deoptimizations
in just the ray-tracer in isolation in \autoref{fig:volcano-comparison-2}. Each
experiment is run 3 times, with 10 iterations and a phase change at iteration 5. In the
first two graphs we changed the type of the height map, in the last one the
numerical interpolation. We observe that deoptless consistently
alleviates the slowdown caused by deoptimization. The first benchmark is simplified
and we manually inlined a numerical computation. In the full version,
changing the type of the height-map produces slightly
slower code in the long run, due to missed optimization
opportunities in the continuation, given that our implementation is still work
in progress.

\begin{figure}[t]
  \includegraphics[width=0.7\columnwidth]{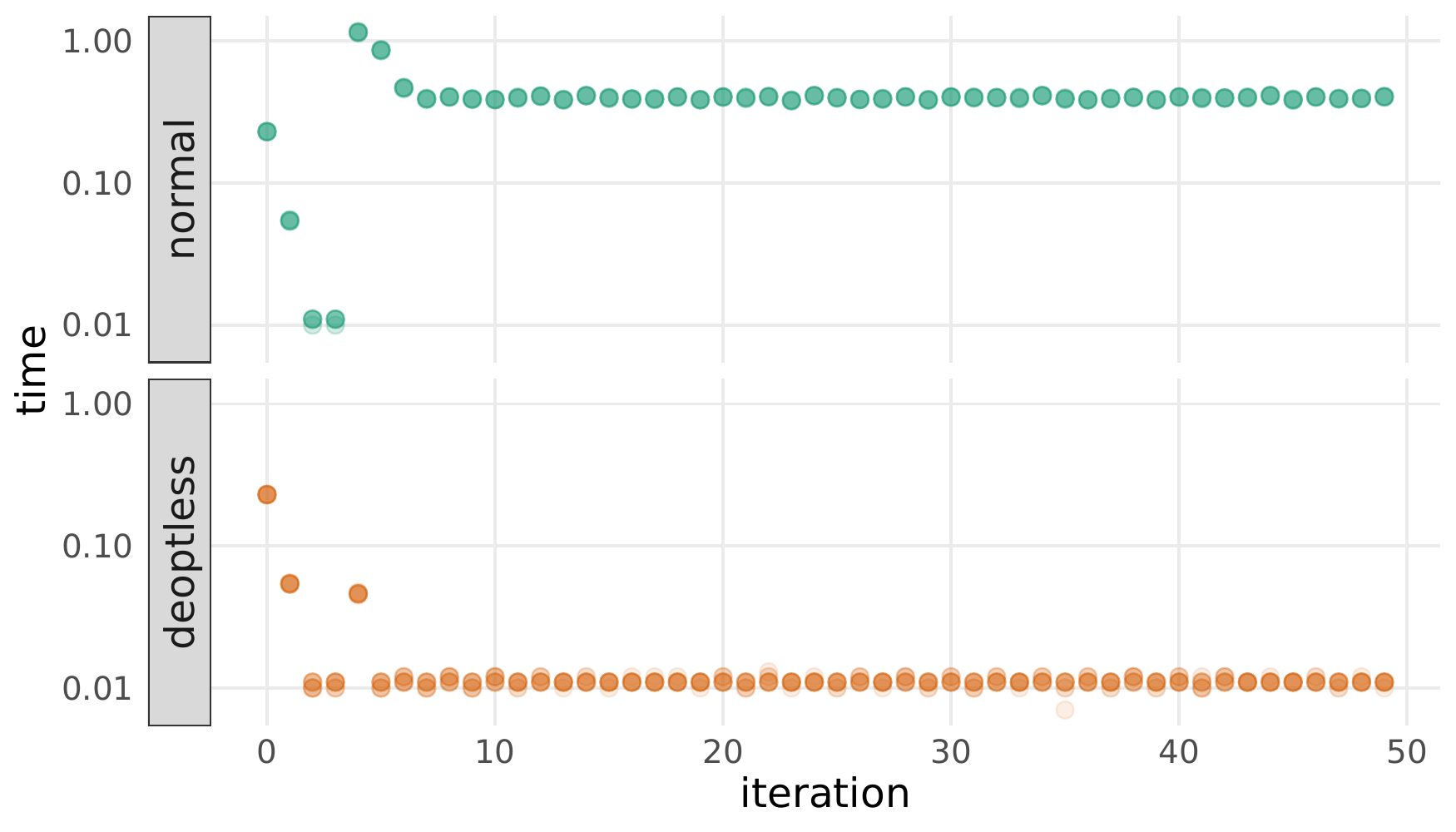}
  \caption{Column wise sum over a table (log scale)}
  \Description{Graph that shows less severe slowdowns with deoptless.}
  \label{fig:bigdataframe-comparison}

  \bigskip
  \ifpreprint\bigskip\else\fi

\begin{lstlisting}[label={lst:bigdataframe-functions},caption={Column-wise sum of a table},style={R}]
    f <- function(colIndex, t) {
        dataCol <- t[[colIndex]]
        res <- 0
        for (i in 1:length(dataCol))
            res <- res + dataCol[[i]]
        res
    }
    columnwiseSum  <- function(t) {
        res <- c()
        for (i in 1L:cols) res[[i]] <- f(i, t)
        res
    }
\end{lstlisting}
\end{figure}

\paragraph{Colsum}

To show how much of an effect deoptless can have in real-world situations, we
investigate the following function in
\autoref{lst:bigdataframe-functions} to summarize the columns of a
table.
We run the benchmark on $\bigDataframeNumberOfColumns$ columns and
$\bigDataframeNumberOfRows$ rows each, consisting of alternating floating-point
and integer columns.
We run and record the run times of \c{f} over $\bigDataframeNumberOfExecutions$ executions
and compare the performance behavior with and without deoptless as shown in \autoref{fig:bigdataframe-comparison}.
In the normal case, the first two iterations include
warmup time spent in the interpreter.
The peak performance reached here is
$\bigDataframeBaselineTimePeakPerformanceSeconds$ seconds.
Then, the fifth iteration corresponds to a float column; a deoptimization
is triggered and control is yielded back to the interpreter where new type feedback is collected
while the function completes. In the deoptless case we see only a temporary
slowdown for compiling the continuation of
$\bigDataframeDeoptlessTimeExecutionAndCompilationDropToSeconds$ seconds.
Considering the stable iterations, deoptless
shows a significant \bigDataframeDeoptlessImprovementAgainstBaseline  performance improvement
over baseline.

\paragraph{Versus Profile-Driven Reoptimization}

\begin{figure}
  \includegraphics[height=1.3in]{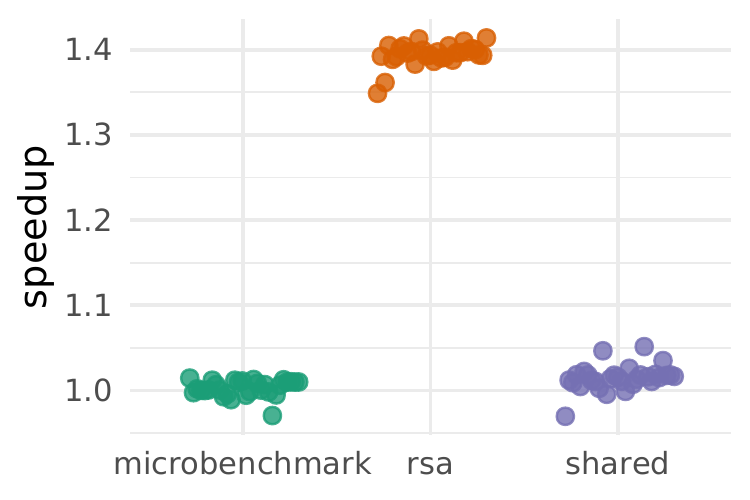}
  \caption{Speedup on reoptimization benchmarks}
  \Description{Graph that shows individual speedup for RSA and same performance
  for the other two.}
  \label{fig:profiler1}
\end{figure}

Finally, we compare the performance profile of deoptless with a
profile-driven reoptimization strategy for \R \citep{dls20}.
The corresponding paper contributes three benchmarks which exhibit problematic cases for
dynamically optimizing compilers. First, a
microbenchmark for stale type-feedback. Then, an RSA implementation, where a key parameter
changes its type, triggering a deoptimization and a subsequent more generic
reoptimization. Finally, a benchmark where a function is shared by multiple
callers and thus merges unrelated type-feedback.
For the three benchmarks they report on speedups of up to 1.2$\times$,
1.4$\times$, and 1.5$\times$ respectively. For deoptless, we expect to
improve only on RSA. In the other two cases the phase change is not accompanied by a
deoptimization, therefore there is no chance for deoptless to improve
performance. We ran these benchmarks against our deoptless implementation
with 3 invocations and 30 iterations;
\autoref{fig:profiler1} presents the results. Each dot
represents the relative speedup of deoptless, for
one iteration of the benchmark each. As expected, the microbenchmark
and the shared function benchmark are unchanged.
The RSA benchmark is sped up by the same amount as in the best case of profile-driven
recompilation.

\section{Conclusion}\label{sec:conclusion}

Speculative optimizations are key for the performance of just-in-time compilers.
They are typically realized using on-stack replacement (OSR) to swap
invalid optimized functions with unoptimized ones on the fly.
In this paper we present a way of dealing with failing speculations
that does not tier down, \ie does not have to continue in a slower tier.
Instead, we bail out of the failed speculation into optimized code. Furthermore, we
use OSR exit points as dispatch points to support multiple specialized
continuations. Thus, instead of having functions become gradually more
and more generic on every deoptimization, we take this opportunity for
splitting and compile functions which become more and more specialized. We present a proof of concept
implementation for \R, an optimizing compiler for the R language. Our
preliminary evaluation shows the big potential of the technique. When presented
with randomly failing assumptions, deoptless is able to execute benchmarks up to $\perfMax\times$ faster than with normal deoptimization, with most
benchmarks being at least $\perfMed\times$ faster and none slower. We also show that
deoptless can improve the peak performance in a number of programs.

As with every forward escape strategy, there is a danger of committing follow-up
mistakes. Deoptless struggles with cases where it is hard to infer from the
failing speculation how the remainder of the function will be affected, before
actually running it. We approach this problem by incorporating information from the current state
of the execution at the OSR exit point. Additionally, we use type-inference on the type-feedback
to override stale profile data. Our evaluation shows that this strategy is
robust and able to produce good code for the continuations.

An interesting avenue for future work would be to try and recombine
continuations into one function again. The information from the contexts
could be used
to recompile and thus get rid of dispatching, as well as code-size overhead, by
fusing everything into one optimized function.

In conclusion, when it comes to speculative optimizations, every mistake is an opportunity to
learn something new. This is certainly true but not helpful, as users do
not wish to wait for their program to learn. For a contemporary approach,
instead of taking a step back and re-analyzing the situation, we
show how to immediately correct our mistakes on the fly, pretend they
never happened, and get away with it.

\begin{acks}
  This work has received funding from the
\grantsponsor{NSF}{National Science Foundation}{} awards
\grantnum{NSF}{1759736}, \grantnum{NSF}{1544542},
\grantnum{NSF}{1925644}, and \grantnum{NSF}{1910850}, the
\grantsponsor{BC}{Czech Ministry of Education, Youth and Sports from
  the Czech Operational Programme Research, Development, and
  Education}{}, under grant agreement No.
\grantnum{BC}{CZ.02.1.01/0.0/0.0/\-15\_003/0000421}, and the
\grantsponsor{ELE}{European Research Council (ERC) under the European
  Union's Horizon 2020 research and innovation programme}{}, under
grant agreement No.  \grantnum{ELE}{695412}.
\end{acks}

\ifpreprint\bigskip\bigskip\else\fi

\bibliography{bib/pir.bib,bib/jv.bib}

\end{document}